# Increased intermolecular interactions and cluster formation at the onset of the twist-bend nematic phase in thioether cyanobiphenyl-based liquid crystal dimers

*Katarzyna Merkel,[1‡] Barbara Loska,[1] Yuki Arakawa,[2] Georg H. Mehl[3], Jakub Karcz [4], Antoni Kocot,[1‡]\**

[1]Institute of Materials Engineering, Faculty of Science and Technology, University of Silesia, ul. 75 Pułku Piechoty, Chorzów 41-500, Poland

[2] Department of Applied Chemistry and Life Science, Graduate School of Engineering, Toyohashi University of Technology, Toyohashi, 441-8580, Japan

[3] Department of Chemistry, University of Hull, Hull HU6 7RX, UK

[4] Faculty of Advanced Technologies and Chemistry, Military University of Technology, Warszawa, Poland

**Abstract**

Infrared spectroscopy (IR) and quantum chemistry calculations, based on density functional theory (DFT) were used to study the structure and the molecular interactions in the nematic (N) and twist-bend ($N_{TB}$) phases of thioether-linked dimers. Infrared absorbance measurements were conducted in a polarized beam for a homogeneously aligned samples to measure and understand the orientation of the vibrational transition dipole moments in the liquid crystal states.

Revealing temperature dependent changes in the mean IR absorbance in the twist-bend nematic phase were observed. In the transition from the N to the $N_{TB}$ phase, is was found to be associated with a decrease in absorbance for the longitudinal dipoles. This is a result of the antiparallel axial interactions of the dipoles, while the absorbance of the transverse dipoles remained unchanged up to 340 K and then increased and the dipoles correlated in parallel. In order to account for the molecular arrangement in the nematic phase, DFT calculations were conducted for the system nearest to the lateral neighborhoods. Changes in the square of transition dipoles were found to be quite similar to the mean absorbance that was observed in the IR spectra. Interactions of molecules that were dominated by pair formation was observed as well as a shift of the long axes of the molecules relative to each other.

The most important observation from the spectroscopic measurements was the sudden increase in the intermolecular interactions as the temperature decreased from the N to the $N_{TB}$ phase. This was evidenced by a significant increase in the correlation of the dipoles that were induced in the dimer cores.

1. **Introduction**

One of the main goals of soft matter theory is to understand the structure-property relationship of material systems in order that the specific properties of the materials can be obtained through molecular design. Self-organization is the process by which molecules spontaneously arrange themselves into stable and ordered structures as a result of non-covalent interactions. This processes can occur on all length scales ranging from the nanoscopic involving atoms or molecules to cosmic-sized objects. These spontaneous organizations contain coded information about the assembling units, i.e., their shape, charge, polarizability, magnetic dipole, mass, etc., as these properties determine the interactions between them [1]. Molecular self-assembly studies provide fundamental information about the influence of intermolecular interactions on the structure of the investigated molecular system. Therefore, such studies are essential in all cases where anisotropic molecular systems can lead to direction-dependent physical properties.

One of the most fascinating examples of molecular self-organization is the emergence of chirality. The formation of chiral superstructures and phases is an area of great interest from both theoretical and practical perspectives [2]. In the biological context, chiral helices are found in DNA and proteins; twisted beta sheets form helical columns [3] that are similar to silk and modern synthetic analogs have also reported [4,5]. In materials science, examples include cholesteric liquid crystalline structures [6] and chiral smectic phases [7], which are formed by bent-core molecules. The most famous example of the last decade of creating chiral structures using non-chiral molecules is the twist-bend-modulated nematic phase ($N_{TB}$), which has a helical structure with a pitch length of several nanometers [8-12]. The structure of the $N_{TB}$ phase has mainly been studied with non-resonant (SAXS, WAXS) [13-15] and resonant X-ray scattering [10,11,16-19], freeze-fracture transmission electron and atomic force microscopy (FFTEM, AFM) [8,9,20-22] polarized Raman [23], infrared [24-26] and nuclear magnetic resonance spectroscopy (NMR) [27-31] as well as with optical methods under applied electric fields [32-38] and magnetic fields [39]. In order to explain the formation of the $N_{tb}$ structure, several models based on rigid cores have been used [15,40,41]. Some models [15, 41] assume the nanophase segregation of the flexible central alkyl linker, terminal chains and the mesogenic cores; molecular end groups attain entropic freedom by associating with the flexible central spacers. This, together with the X-ray observation of the half-molecular length periodicity along the $N_{TB}$ helix, led to a proposal of a model of the self-assembly of half molecule-long segments into helical tiled chains of molecules as the basic structural element of the $N_{TB}$ phase [10,18,42]. The $N_{TB}$ has been most investigated cyanobiphenyl-based liquid crystal dimers (CBCnCB, n = 7.9.11) [8-10,11,13,16,21,23,27-33,36-39] where the cyanobiphenyl units tend to self-associate into antiparallel pairs *via* dipole-dipole interactions, [43-46]. The extensive research on the structure of the $N_{TB}$ phase, to date, permits specific conclusions to be drawn, namely, that a molecular curvature is fundamental for the formation of the phase and that the stability of this phase increases with a decrease in the molecular bending angle [47-52]. There have also been important reports suggesting that the N-$N_{TB}$ phase transition is also influenced by such factors as: intramolecular torsion [17,53,54], conformational changes [55-59], bend angle fluctuations [60], the effect of free volume [55, 61,62] as well as intermolecular interactions [25,63-67].

It has long been known that hydrogen bonds such as the $\pi - \pi$ intermolecular interactions play a very important role in the supramolecular biological phenomena such as base-pairing in double stranded DNA, protein binding, cell-cell recognition and viral infection [68,69]. A similar phenomenon is observed in the field of soft matter where unconventional intermolecular interactions lead to the emergence of new supramolecular liquid crystal structures [70,71] such as the twist-bend phase [72] or the polar-twisted phase ($N_{TB}$) [73] as well as the newly discovered ferroelectric nematic phase ($N_F$) [74,75].

This work is devoted to observations of the intermolecular interactions and their impact the structural changes that occur on the transition from the nematic phase to the twist-bend phase. For this purpose, infrared spectroscopy, which is an excellent tool for analyzing rotational order in a molecular systems was used. In the case of conventional nematics, it is usually assumed that the transition dipoles that correspond to a specific vibration are temperature independent. As absorbance is related to the square of the transition dipole, the average absorbance is expected to be density dependent. Density in turn is typically temperature dependent. In the $N_{TB}$ phase, however, the average absorbance of a number of molecular vibrations exhibits a specific behavior and this is due to the self-assembly of the molecular system. Experimental data indicate that short-range interactions grow significantly and thus the inter-correlations between the transition dipoles of the interacting functional groups become more important. Also the behavior of the longitudinal dipoles are different to those of the transversal dipoles.

Here, we report on FTIR and DFT studies for two groups of dimers: a symmetric system containing the thioether-linking groups (C-S-C) with the acronym CBSCnSCB (n=5,7) and a materials containing ether-linking groups (C-O-C) with the acronym CBOCnOCB (n=7) and an asymmetric compound that contains both the ether- and thioether-linking groups, termed CBSCnOCB (n=5,7).

2. Materials and Methods

2.1. Materials

Symmetrical and asymmetrical liquid crystal dimers with the cyanobiphenyl (CB) mesogenic groups were investigated. We present three symmetric dimers with the general acronym CBXC7XCB (X=C or S or O), which contain nine functional groups in the chain that links the two mesogenic cores (methylene, thioether and ether): the system CBC9CB, which contains the nine methylene groups in the linker and CBSC7SCB and CBOC7OCB in which the alkyl chains that contain seven methylene groups are connected to the cyanobiphenyls by two thioether or two ether bridges, respectively. In the asymmetric dimers with the acronym CBSCnOCB (n=5,7), the mesogens were linked to an alkyl chain on one side with five or seven methylene groups by a thioether bridge and on the other by an ether bridge. The basic CBC9CB compound was synthesized as described in Ref. [76-79]. All of the thioether/ether compounds were synthesized in the Department of Applied Chemistry and Life Science at the Toyohashi University of Technology (Japan) and the details of the synthesis are presented in the papers [80,81].

## 2.2. Infrared spectroscopy

The planarly aligned cells were prepared between two optically polished zinc selenide (ZnSe) discs. The thickness of the fabricated cells was determined to be in the range of 5.1-5.6 μm by measuring the interference fringes using a spectrometer that was interfaced with a PC (Avaspec-2048). The infrared spectra were acquired using a Fourier infrared spectrometer (Agilent Cary 670 FTIR). The experiment was conducted using the transmission method with a polarized IR beam. An IR-KRS5 grid polarizer was used to polarize the IR beam. The IR spectra were measured as a function of the polarizer rotation angle in the wavenumber range 500-4000 $cm^{-1}$. Details of sample preparation and the absorbance measurements were reported earlier [26]. These measurements enabled the orientation of the transition dipole moment of the bands to be determined with respect to the long molecular axis as well as the temperature dependencies of the absorbance of the samples. To determine all three components of absorbance ($A_X$, $A_Y$ and $A_Z$), it was necessary to measure two samples with different orientations: planar (homogeneous) and homeotropic. Unfortunately, in the case of the tested materials, i.e., for the cyanobiphenyl dimers, it was extremely difficult to obtain a good homeotropic alignment. Therefore, in order to calculate the mean absorbance of the sample, assuming that the material was uniaxial, it was assumed that $A_X = A_Y$ and the mean absorbance was determined as $A_0=(2A_X+A_Z)/3$. The absorbance components were determined as being the area that was bound by the contour of the given band using Bio-Rad Win-IR Pro version 2.96e. In the case of complex bands that contained more vibrations, they were separated using the Origin Pro 2021 software using the Pearson VII fit. Figure 1 shows the configuration of the infrared measurements using the polarized transmission technique (Fig. 1a) and the molecular structure of the cyanobiphenyl dimers (Fig. 1b).

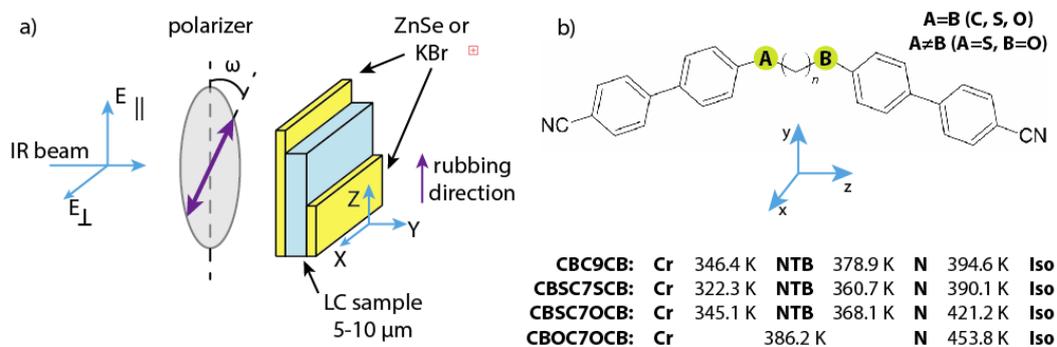

Figure 1. a) Schematic of the polarized infrared transmission technique at a normal incidence of light. The laboratory frame (X,Y,Z) for the planar cell. In the nematic phase, Z was the axis along and X was the axis perpendicular to the optical axis (the optical axis coincided with the rubbing direction). In the $N_{TB}$ phase, Z coincided with the helix axis. b) Transition temperature and molecular structure of the cyanobiphenyl dimers with a molecular frame of reference: z – long axis (bowstring), x – axis normal to the bent plane, y – bow arrow axis.

*2*.3. Intermolecular non-specific binding energy in the system – DFT study

In this work, the calculations of the electronic structure of the molecules were performed using the Gaussian09 software package (version E.01) [82]. The molecular structures, intermolecular binding energy, harmonic vibrational force constants, absolute IR intensities and components of the transition dipole moment were calculated using the density functional theory (DFT) with the Becke's three-parameter exchange functional in combination with the Lee, Yang and Parr correlation functional B3-LYP method with the basis set: 6-311(d,p) [83]. The results were visualized using GaussView 6.

From the point of view of vibrations of individual molecular groups, we practically do not observe the coupling between vibrations for the rigid mesogen cores (arms) of the same dimer [24-26]. On the other hand, the interactions of arms (molecular groups) belonging to adjacent dimers turn out to be significant [25-26]. Therefore, in order to confirm the experimental observation and to determine the molecular arrangement in the $N_{TB}$ phase, DFT calculations were performed for the system including the nearest the lateral neighborhoods. For the reasons described above as well as because DFT calculations for such large molecules are very time consuming, calculations were made for the CBSC7, the CBOC7 monomers. First, optimization of the isolated molecules of all dimers/monomers was performed and potentially stable conformers were determined based on the calculation of the energy barriers for the internal rotation of the cyanobiphenyl and the rotation around the dihedral angle between the cyanobiphenyl and the linker. All details were described and discussed in a separate paper [84].

In the next step, the six monomer molecules (the optimized molecule and the most energetically stable conformer were used [84]) were arranged in the so-called a sublayer (Fig. 1a). The arrangement of the molecules in such a sublayer was created based on the experimental data from the resonant X-ray scattering measurements (TReXS) [12].

We prepared two systems: one that contains only the same CBSC7 molecules and the mixed one, with three CBSC7 and three CBOC7 molecules. Figure 2a shows, for example, a top view of the initial arragnment CBSC7 molecule for calculation. Molecules were arranged parallel to each other and the distance from them was approx. 5Å and their cyano groups were arranged alternately (Fig. 2b). Such a system was optimized.

In the next stage, the atoms and bonds of the molecules that were located at the periphery of the system were frozen. The frequencies and intensities of the vibrations were calculated for a central molecule that was surrounded by other molecules (Fig.2a, the selected molecule highlighted in blue). This approach allowed the calculation of the FTIR spectrum, and more precisely the square of the dipole transition moment for a single molecule, but taking into account the influence of the nearest surroundings.

The interaction energy for the system was determined using the so-called supermolecular approach using the base superposition error (BSSE) correction utilizing the counterpoise method (CP). The intermolecular binding energy was estimated at 30 and 37 kJ / mol (for the thioether and mixed system respectively).

$$\Delta E^{CP} = E^{AB} - E^{A(AB)} - E^{B(AB)} \qquad (2)$$

where $E^{AB}$ is the energy of the complex, $E^{A(AB)}$ – the energy of fragment A, which was calculated in the dimer base and, $E^{B(AB)}$ – the energy of fragment B, which was calculated in the base of the dimer. Hence BSSE was equal to:

$$BSSE = E^{A(AB)} - E^{B(AB)} = E^A - E^B \qquad (3)$$

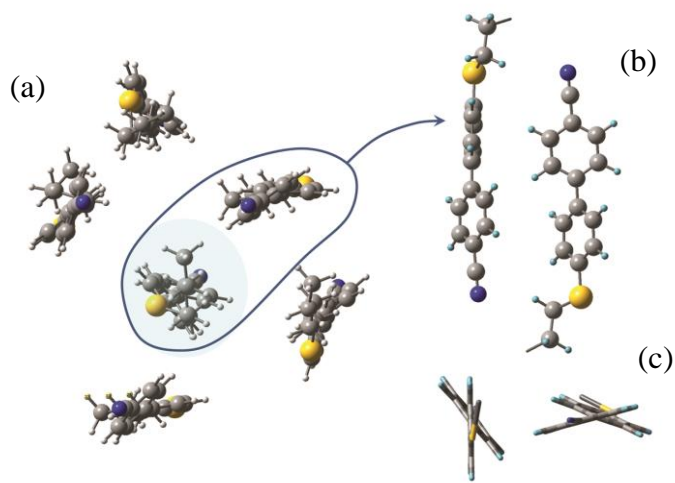

Figure 2. Side arrangement of the CBSC7 monomer parts. a) Top view (X-Y plane) of the system. Initial arrangement of six molecules into a sublayer prepared for DFT calculations. IR spectra were calculated for a molecule (highlighted in blue) that was surrounded by other molecules. b) The close-up view of the CBSC7 molecules pair formed by intermolecular π-π interaction of cyanobiphenyl after optimization using B3LYP/6-311+G method. d) Top view (X-Y plane) of a pair of molecules.

3. **Results and discussion**

   3.1. Temperature dependencies of the absorbance

By measuring the absorbance in polarized light, the temperature dependencies of the absorbance components in the temperature range of the N and the $N_{TB}$ phases could be directly analyzed [24-26]. We expected a distinctly different behavior of the absorbance components for the bands that had a transition dipole that was longitudinal and transverse with respect to the core axis.

The average intensity of absorbance, $A_0$, and the related transition dipoles were analyzed for all of the dimers in the temperature range of the N and the $N_{TB}$ phases. In the range of the N phase, the average absorbance increased as the number density of the molecules increased, which incidentally confirmed that the transition dipole appeared to be constant in this temperature range. In the temperature range below the $N_{TB}$ transition, the behavior was different for the symmetric and asymmetric dimers and in some cases was dependent on the substrate of the cell windows [26].

A detailed analysis of the infrared spectra compared to those simulated for different conformers, along with the exact assignments of vibrations, is presented in a separate paper [83]. Several vibrational bands that belonged to the longitudinal and transverse transition dipole that offered a significant dichroism of the band were selected to be analyzed. For the longitudinal transition dipole, the phenyl stretching band ($\nu CC$) at 1600 cm$^{-1}$ and the combinational band at 1100 cm$^{-1}$ ($\nu_{as}C_{Ar}S + \beta CH\ ip\ CB$) were selected. Figure 3 shows the temperature dependencies of the average absorbances, $A_0$, for the bands that corresponded to the longitudinal transition dipole. There was a clear transition from the conventional N phase to the $N_{TB}$ for all of the molecules except for CBOC7OCB, which behaves like a classical calamitic molecule, due to its large opening angle and does not form the $N_{TB}$ phase. In contrast, the absorbance band that corresponds to the longitudinal transition dipole at the 1600 cm$^{-1}$ (or 2300 cm$^{-1}$, assigned to the stretching vibration of the cyano group, $\nu CN$) underwent a significant decrease just after entering the $N_{TB}$ phase (Fig. 3). This behavior clearly indicated that the longitudinal dipoles of the neighboring cyanobiphenyl groups show a tendency for antiparallel arrangements. The strong dipole of the CN group induces a dipole in the adjacent biphenyl molecule. As a result, intermolecular van der Waals interactions were observed (more precisely, the London dispersion forces), which leads to the stabilization of the $N_{TB}$ phase. It was expected that a nano-segregation process might occur along the Z-axis of the system. This behavior has already been reported for cyanobiphenyl LCs, showing an overlap of the termini of the molecules [44-46,62]. Following the transition to the $N_{tb}$ a significant decrease in the average absorbance was measured, suggesting the presence of dipole correlations in this phase and the presence of additional molecular interactions that are related to the change in the geometry of the system.

We also noticed that in the case of the CBSC7SCB dimer, there was a somewhat smoother decrease in the mean absorbance for the band at 1100 cm$^{-1}$ than for the band at 1600 cm$^{-1}$. The maximum at the 1100 cm$^{-1}$ involves a sulfide bridge as part of the linker and therefore describes the direction of the long axis of the dimer more than the band at 1600 cm$^{-1}$. In the case of the cyanobiphenyl bands (2200, 1600, 1485 cm$^{-1}$), the transition dipole was significantly inclined from the long molecular axis as it was found that the vibrations of its counterparts in the dimer were almost decoupled from each other. This phenomenon was already discussed previously [26] when determining the order parameters.

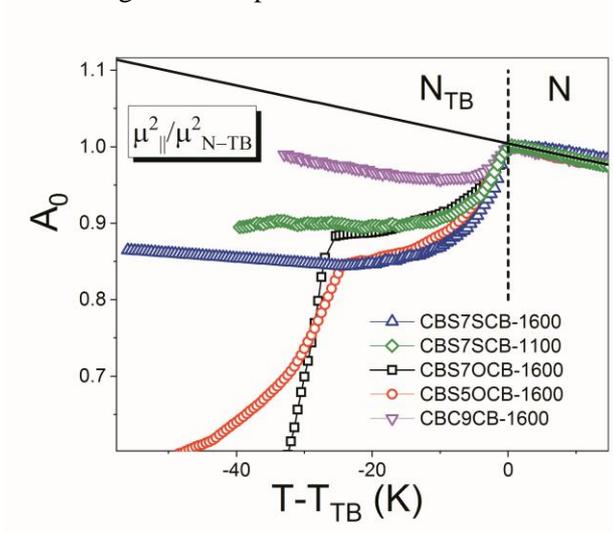

Figure 3. The average absorbance vs. temperature behavior of the dimers for the longitudinal dipole ($\mu_\parallel$): benzene ring vibration at 1600 cm$^{-1}$ ($\nu$CC) and a deformation vibration of the C-H group in the benzene ring plane at 1100 cm$^{-1}$ ($\beta$CH ip CB + $\nu_{as}C_{Ar}S$). □ – CBSC7OCB, ○ – CBSC5OCB, △ – CBSC7SCB (1600 cm$^{-1}$), ▽ – CBC9CB; ◇ –CBSC7SCB (1100 cm$^{-1}$); *solid black line* – nematic phase trend line

Three bands were selected for the transverse transition dipole: the out of plane deformation vibrations of the C-H groups at 811 cm$^{-1}$ ($\gamma$CH op CB), the complex band at 821 cm$^{-1}$ ($\gamma$CH op CB + $\nu_s$ C$_{Ar}$O; s- symmetric) and the in-plane deformation of the C-H groups at 1395 cm$^{-1}$ ($\beta$CH ip CB with a sulfur linkage C-S-C). Figure 4 shows the temperature dependencies of the average absorbances, $A_0$, for the transverse transition dipole. The average absorbance of the transversal dipole (811 cm$^{-1}$) maintained its trend until the crystallization temperature was reached. For the asymmetric dimers (CBSCnOCB), two regions of the temperature range of the N$_{TB}$ phase could clearly be distinguished: the first extending more than 20 K below the N-N$_{TB}$ transition temperature and the next below ~340 K and almost approaching room temperature. In the first region, absorbances of the transversal dipoles maintained the trend due to an increase in the number density of molecules until the temperature reached 340 K, which was similar to the results observed for CBSC7SCB.

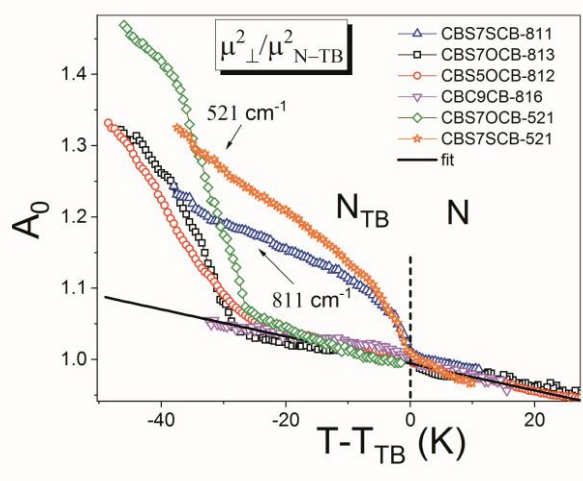

Figure 4. The average absorbance vs. temperature behavior of the dimers for the transversal dipole ($\mu_\perp$): 811 (813/816) cm$^{-1}$ ($\gamma$CH op CB for the rigid core with a sulfur linkage C-S-C): □ – CBSC7OCB, ○ – CBSC5OCB, △ – CBSC7SCB, ▽ – CBC9CB; 520 cm$^{-1}$ ($\gamma$CC op CB + $\delta$CS): ◇ – CBSC7OCB, ★ – CBSC7SCB.

By contrast, in the lower temperature region below 340 K, the absorbance departed from the trend that was predicted by the density number of molecules, this was accompanied by further absorbance intensity drop of the longitudinal dipole. Such behavior of the absorbance of the transversal dipole might indicate that some new intermolecular interactions has led to the orientation of the bond. It seems that the sulfur group was either involved in the London dispersion forces, π-stacking interactions, or in specific hydrogen bonds with neighboring molecules. It is worth noting that the vibration that involved the sulfur linkage (~811 cm-1) showed an increase in absorbance while that of the oxygen linkage (~820 cm-1) decreased in absorbance intensity. Such a difference in the behavior of the transverse dipoles suggests that the phase symmetry is no longer uniaxial. Moreover, the most probable conformations of the dimers changed and became specific due to a possible rotation around the sulfur bridge.

Considering the high rotational barrier around the bond connecting the cyanobiphenyl group to the alkyl linker, which is much higher in the case of the oxygen bridge than for the sulfur bridge [12,83], we expected that the oxygen bridge with the phenyl group would remain in the z-y plane. In contrast, the sulfur bridge might rotate by a significant angle with respect to the plane of the phenyl group. If we assume that the absorbance changes at 340 K were due to an order rearrangement (phase reorganization), we might consider a phase biaxiality.

These findings motivated us to analyze the possible intermolecular interactions that could lead to the bond orientation. In such a system, it might be possible to select a specific type of interaction that might change the transient dipole moment of the vibrating bands. We also used DFT modeling to analyze the possible intermolecular interactions that would lead to the bond orientation and thus the stabilization of the twist-band phase.

### 3.2. Hypothetical arrangement of the CBSC7SCB molecules that were formed by the intermolecular weak interactions – DFT modeling

In order to confirm our assumptions about the geometry of the helix and its close-packed structures in which weak intermolecular interactions might play a significant role, the geometry of the system was optimized with the six interacting molecules (monomers with a sulfur atom CBSC7, Fig. 2a, and the mixed system containing the CBSC7 and the CBOC7 molecules,). The details and stages of the performed simulations were described in paragraph 2.3. After optimization, the distances of the molecules changed significantly and ranged from 3.1 to a maximum of 3.9 Å, the initial distance was about 5 Å. A grouping of the molecules into pairs as well as a shift of the molecules' axes relative to each other was detected (see. Fig 2b). As observed in numerous physical experiments a nano-segregation into pseudo-layers – biphenyls "stacking" next to each other creating an aromatic pseudo-layer and the spacer tending to be parallel in order to form an aliphatic layer was observed also. For molecules with a sulfide bridge, the molecules aligned with each other in such a way that the phenyl planes of cyanobiphenyls remained in the so-called "T-shape" conformation (Fig.2c) relative to neighbors. In the case of oxygen bridge molecules, a flatter "parallel-displaced" arrangement was observed. The segregation of the aromatic parts (partial overlapping) as well as an attraction of the CN group to the alkyl chains was observed. Molecular interdigitation also occurred too [44,46].

In the next stage, the outer molecules were frozen, while the vibrational frequencies and transition dipole moments were calculated for the molecule that remained in the center (see. Fig 2a). The theoretical IR spectrum for a surrounded molecule should reflect the situation for which the intermolecular forces (IMF) are taken into account. The IR spectrum of a molecule where the intermolecular forces were taken into account was compared with the spectrum of the isolated molecule as well as with the experimental spectrum in the N and $N_{TB}$ phases (Fig. 5 & Fig.6). Figure 5 shows the comparison of the theoretical spectra of the CBSC7 molecule for the system with intermolecular forces (IMF) with the situation without IMF (case of isolated molecule) and the experimental spectra. Figure 6 shows the comparison of the theoretical spectra of the CBOC7 molecule in the mixed system with intermolecular forces (IMF) with spectra of isolated molecule and the experimental spectra. For the longitudinal dipoles, (bands at wavenumbers 1000, 1100, 1485, 1600 and 2300 $cm^{-1}$), a decrease in the intensity of the bands for the system with the IMF when compared to a single monomer was detected, a result consistent with the correlation along the z-direction, visible in the IR experiments. An increase in the intensity for the transverse bands was also observed, i.e., for the bands at 520 and 811 cm$^{-1}$. This result implies a correlation of the dipoles in this direction.

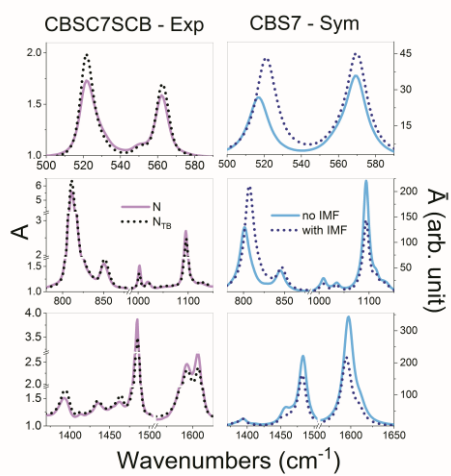

Figure 5. Comparison of the theoretical spectra for the system with intermolecular forces (IMF) without intermolecular forces and the experimental spectra. Simulations for the CBS7 monomer: *solid blue line* – spectra for an isolated molecule (no IMF), *short navy dotted line* – system of six interacting molecules (with IMF). Experimental spectra of the CBS7SCB dimer, *solid magenta line* – the nematic phase, *short black dotted line* – the twist-bend nematic phase.

For the bands with the transversal dipoles (520, 811 cm$^{-1}$), an increase in the intensity of the bands by approximately 50% relative to a isolate molecule was detected and for the vibrations in the longitudinal dipoles (1100, 1485, 1600 cm$^{-1}$), the decrease was approximately 40%. The smallest change in intensity was obtained for the CN vibrations at 2300 cm$^{-1}$, by approximately 10%). These values are in line with the trend that was observed in the experimental spectra at the transition from the N to the N$_{TB}$ phase. Table 1 summarizes the changes in the dipole moment of the transition: $\frac{\mu_{TB}^2}{\mu_N^2}$ in the presence of weak intermolecular interactions and local orientational order.

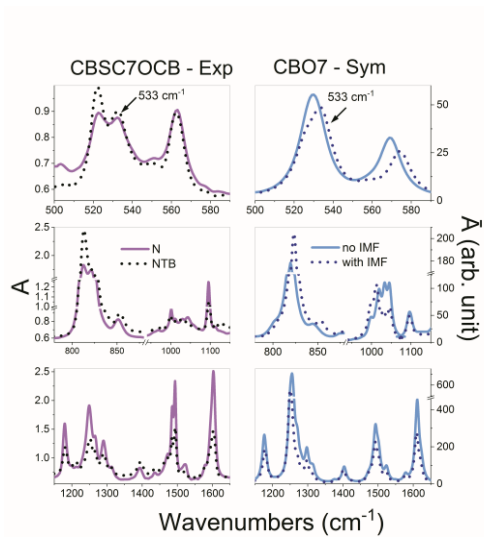

Figure 6. Comparison of the theoretical spectra for the system with intermolecular forces (IMF) without the intermolecular forces and experimental spectra. Simulations for the CBO7 monomer: *solid blue line* – spectra for the isolated molecule (no IMF), *short navy dotted line* – system of six interacting molecules (with IMF). Experimental spectra of the CBS7OCB dimer: *solid magenta line* – the nematic phase, *short black dotted line* – the twist-bend nematic phase.

Table 1. Changes of the transition dipole moment: $\frac{\mu_{TB}^2}{\mu_N^2}$ in the presence of weak intermolecular interactions and local orientational order.

∗ DFT simulation – B3LYP/6-311G (p,d)

| Vibration frequency (cm$^{-1)}$ | $\mu_{TB}^2/\mu_n^2$ and ∗$\mu_{IMF}^2/\mu_{noIMF}^2$ | | | | | |
|---|---|---|---|---|---|---|
| | CBSC7SCB | CBSC7SCB DFT* | CBC5OCB | CBC7OCB | CBC7OCB DFT* | CBC9CB |
| 520 | 1.25 | 1.60 | 1.42 | 1.48 | 1.54 | -- |
| 810 | 1.22 | 1.64 | 1.32 | 1.33 | 1.41 | 1.05 |
| 1100 | 0.94 | 0.65 | 0.94 | 0.84 | 0.84 | -- |
| 1250 | -- | -- | 0.89 | 0.71 | 0.80 | -- |
| 1485 | 0.89 | 0.73 | 0.90 | 0.66 | 0.70 | 0.94 |
| 1600 | 0.86 | 0.63 | 0.78 | 0.60 | 0.61 | 0.93 |
| 2220 | 0.80 | 0.93 | 0.85 | 0.68 | 0.95 | 0.93 |

We realize that the interactions between several molecules do not fully explain the behavior of a group of molecules. Thermodynamic and probabilistic concepts such as entropy, enthalpy and free energy should be taken into account to determine this. Hence our current calculation are the first step rowardsmore advanced simulations using periodic density functional theory modeling, which we are planning to perform in the future.

Based on the analysis of the molecular structure, a  model of the geometrical arrangement of the molecules in the $N_{TB}$ phase was developed (Fig. 7). In the model, the molecules follow the shape of the helix formed by the director, with overlapping biphenyl groups linked to increased bond/dipole correlations.  as was shown by the optimization of the geometry of the monomer system. The detailed nature of the effects stabilizing the $N_{TB}$ phase as still not fully  understood, For the investigated systems the hydrogen bonds between the CN group and the hydrogen atoms of thespacer, the hydrogen bonds between the sulfur atom and the hydrogen atom of the benzene ring or the interaction of the π-π orbitals of the benzene rings all play a role.  Nevertheless, the effect of intermolecular interactions in the phase is clearly visible. Both the model and the experimental results for intermolecular interactions are consistent with the reports of other measurements that have been obtained using the TReXS (Tender Resonant X-ray Scattering) method [11,12].

Table 2. Summary of the determined molecular parameters: the length of the molecule ($l$), the tilt angle ($\theta_t$), the helix pitch ($p$), the effective length of the molecule ($d$), the number of molecules per helix pitch ($p/d$).

| CBCnCB | $\theta_t$ (°) | $p$ (nm) | $l$ (nm)* | $d = l\cos\theta_t$ | $p/d$ |
|---|---|---|---|---|---|
| **CBC9CB** | 25.6 | 8.15 [10] | 2.88 | 2.6 | 3.1 |
| **CBSC7SCB** | 33 | 9.1 [11,12] | 2.90 | 2.43 | 3.7 |
| **CBSC7OCB** | 15.6 | 11.5 [12] | 2.90 | 2.79 | 4.1 |

∗ − lengths of the molecules were determined for the optimized molecules using the B3LYP / 6-31 G (d, p) method (CBC9CB − U conf., CBSC7SCB − F conf., CBSC7OCB − F conf., see paper [84]

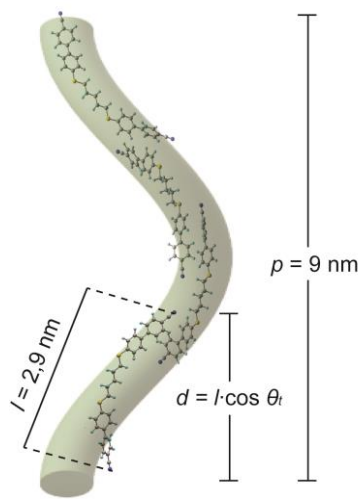

Fig. 7. Proposed hypothetical arrangement of the thioether dimers in the twist-bend nematic phase. The helix pitch was computed as $p = 2\pi / q$ where q is the wave vector.

4. CONCLUSIONS

The most important observation from the spectroscopic measurements was the sudden increase in the intermolecular interactions as the temperature decreased from the nematic phase to the twist-bend phase. This was evidenced by a significant increase in the correlation of the dipoles that were induced in the rigid dimer cores. The longitudinally induced dipoles of cyanobiphenyl had negative correlations (anti-parallel arrangement), while the perpendicular dipoles were positively correlated. To explain this self-assembling dimer pairs were analyzed using DFT calculations and optimizing the energies of their interactions. The results most consistent with the experimental data is a system of several molecules that exhibits a clustering of the rigid cores (cyanobiphenyls) and nonspecific weak intermolecular interactions. These interactions are mainly associated with the interactions of the π-π orbitals of the aromatic rings and the sulfur atoms, however the formation of weak hydrogen bonds H⋯S/O, H⋯N it may additionally also be present. Such non-specific intermolecular interactions led to a significant bond ordering in the chiral phase.

Based on the DFT simulation for the groups of interacting molecules along with the experimental data that was obtained from the absorbance measurements as well as X-ray resonance scattering, a model for the packing of the dimer molecules in the twist-bend phase was developed. The proposed model assumes an overlapping / interpenetration of the rigid cyanobiphenyl cores, which might stabilize the $N_{TB}$ phase and affect the helical pitch.

ACKNOWLEDGEMENTS

Authors K.M. & A.K. thank the National Science Centre for funding through the Grant No. 2018/31/B/ST3/03609. B.L. thanks the National Science Centre for funding through the project No. 2020/39/O/ST5/03460.

All DFT calculations were carried out with the Gaussian09 program using the PL-Grid Infrastructure on the ZEUS and Prometheus cluster.

AUTHOR INFORMATION

**Corresponding Authors:**

* E-mail: antoni.kocot@us.edu.pl  (A.K)

**E-mails co-authors:**

Katarzyna Merkel: katarzyna.merkel@us.edu.pl
Barbara Loska: barbara.loska@us.edu.pl
Yuki Arakawa: arakawa.yuki.xl@tut.jp
Georg Mehl: g.h.mehl@hull.ac.uk
Jakub Karcz: jakub.karcz@wat.edu.pl
**ORCID:**
Antoni Kocot: 0000-0002-9205-449X

Barbara Loska: 0000-0002-0756-5018

Yuki Arakawa: 0000-0002-8944-602X

Georg Mehl: 0000-0002-2421-4869

Jakub Karcz: 0000-0003-4783-0469

Katarzyna Merkel: 0000-0002-1853-0996

**Author Contributions:**
[‡] K.M and A.K contributed equally to this work.

Conceptualization: K.M, A.K
 Synthesis:  G.H.M, J.K,Y.A

Methodology: K.M, A.K
Investigation: K.M, B.L
Writing—original draft: K.M
Writing—review & editing: K.M, A.K

**Funding Sources:**

National Science Centre, Poland for Grant No. 2018/31/B/ST3/03609

National Science Centre, Poland for Grant No. 2020/39/O/ST5/03460

**Competing interests:**

Authors declare that they have no competing interests.